\newcommand{\be}{\begin{equation}}\newcommand{\ee}{\end{equation}}
\newcommand{\bea}{\begin{eqnarray}}\newcommand{\eea}{\end{eqnarray}}
\newcommand{\nn}{\nonumber}
\documentstyle[12pt]{article}
\begin{document}
\renewcommand{\thefootnote}{\fnsymbol{footnote}}
\thispagestyle{empty}
\begin{center}
THE $N=2$ SUPER-$W_3^{(2)}$ ALGEBRA
\vspace{1cm} \\
S. Krivonos${}^{a,b}\;$\footnote{E-mail: krivonos@thsun1.jinr.dubna.su},
A. Sorin${}^{b}\;$\footnote{E-mail: sorin@thsun1.jinr.dubna.su} \vspace{1cm}\\
${}^a$INFN-Laboratori Nazionali di Frascati, P.O.Box 13 I-00044 Frascati,
Italy\\
${}^b$Bogoliubov  Theoretical Laboratory, JINR, Dubna, Head Post Office,\\
P.O.Box 79, 101 000 Moscow, Russia
\vspace{3cm} \\
{\bf Abstract}
\end{center}
    We construct the nonlinear $N=2$ super-$W_3^{(2)}$ algebra  with
an arbitrary central charge at the classical level in the framework of
Polyakov "soldering" procedure.
It contains two non-intersecting subalgebras:
$N=2$ superconformal algebra and $W_3^{(2)}$ and their closure gives the
$N=2$ super-$W_3^{(2)}$ algebra.
Besides the currents of $N=2$ superconformal and $W_3^{(2)}$ algebras, it
comprises two pairs of fermionic currents with spins 1 and 2.
The hybrid field realization and contractions to the zero central charge
are constructed.\vspace{1cm}\\
\begin{center}
{\it Submitted to Phys. Lett. B}
\end{center}
\vfill
\setcounter{page}0
\renewcommand{\thefootnote}{\arabic{footnote}}
\setcounter{footnote}0
\newpage

{\bf 1. Introduction}\\
In the last years the progress was achieved in
understanding  the procedure
of the supersymmetrization of $W$-type nonlinear algebras and some
nonlinear superalgebras were constructed both at the classical and
quantum levels (see, for instance, Ref.\cite{s} and references therein).
All such superalgebras comprise the
currents with canonical conformal spins: integer for bosonic currents and
half-integer  for fermionic ones. However, among the nonlinear bosonic
algebras there exist
special algebras which contain the bosonic currents
with non-canonical half-integer spins [2-6].
Up to now the question of
their supersymmetric extensions is open.

The $W_3^{(2)}$ algebra \cite{p,b} is the simplest nontrivial example of
such algebras.
It represents the bosonic analogue of the two-dimensional chiral
$N=2$ superconformal algebra(SCA) \cite{x} and contains
bosonic currents with noncanonical 3/2 spins.
It is interesting to find
its supersymmetric extensions which would comprise also the fermionic
currents with non-canonical integer spins.
In the present letter we  construct  the $N=2$ supersymmetric extension of
the $W_3^{(2)}$ algebra and its hybrid field realization.

{\bf 2. Preliminaries} \\
By reason of minimality, let us suppose that
the $N=2$ super-$W_3^{(2)}$ algebra, we are looking for,
contains the $W_3^{(2)}$ and $N=2$
SCA as sub-algebras and it does not comprise additional
bosonic currents, besides the ones corresponding to these subalgebras.
To understand their possible embeddings in
the $N=2$ super-$W_3^{(2)}$, let us discuss some
formal relations between the $W_3^{(2)}$ and $N=2$ SCA
which a priory will allow us to make some heuristic conjectures.

    The $W_3^{(2)} \propto \left\{ J_w, G^{+}, G^{-}, T_w \right\}$ and
$N=2$ SCA $\propto \left\{ J_s, S, \bar{S}, T_s\right\} $ contain the
currents with the same spins $\left\{1,3/2,3/2,2\right\}$ respectively,
but while
the currents $ S,\bar{S}$ are fermionic, their counterparts $G^{+},G^{-}$
are bosonic.
The operator-product-expansions (OPE) for these algebras
have the following structure \cite{p,b,x}
\footnote{Hereafter we do not adduce  OPE without singularities.
All currents appearing in the right-hand sides of the OPE are evaluated at
point $z_2$. We strictly fix the correlation between
the central charges of the $W_3^{(2)}$ and $N=2$ SCA,
which will be understood later.}
\be \label{1}
W_3^{(2)} : \left\{
\begin{array}{l}
J_w(z_1)J_w(z_2)  = \frac{\frac{c}{6}}{z_{12}^2}  \quad , \quad
J_w(z_1)T_w(z_2)  =  \frac{1}{z_{12}^2}J_w,   \\
J_w(z_1)G^{\pm}(z_2)  =\mp  \frac{\frac{1}{2}}{z_{12}}G^{\pm} \quad , \quad
T_w(z_1)G^{\pm}(z_2)  = \frac{\frac{3}{2}}{z_{12}^2}G^{\pm}
+ \frac{1}{z_{12}}{G^{\pm}}', \\
T_w(z_1)T_w(z_2)  =  \frac{-3c}{z_{12}^4}+\frac{2}{z_{12}^2}T_w  +
                         \frac{1}{z_{12}}T_w',  \\
G^{+}(z_1) G^{-}(z_2)  =  \frac{2c}{z_{12}^3}-\frac{6}{z_{12}^2}J_w  -
        \frac{1}{z_{12}}(T_w-\frac{12}{c}J_w^2+3J_w'),
\end{array}
\right.
\ee
\be\label{2}
\mbox{N=2 SCA} : \left\{
\begin{array}{l}
J_s(z_1)J_s(z_2)  =  \frac{\frac{c}{2}}{z_{12}^2} \quad , \quad
J_s(z_1)T_s(z_2)  =  \frac{1}{z_{12}^2}J_s,  \\
J_s(z_1)S(z_2)  =  \frac{\frac{1}{2}}{z_{12}}S  \quad , \quad
J_s(z_1)\bar{S}(z_2)  = \frac{-\frac{1}{2}}{z_{12}}\bar{S}, \\
S(z_1) \bar{S}(z_2)  =  \frac{2c}{z_{12}^3}+\frac{2}{z_{12}^2}J_s  +
        \frac{1}{z_{12}}(T_s+J_s'),   \\
T_s(z_1)S(z_2)  =  \frac{\frac{3}{2}}{z_{12}^2}S+
                         \frac{1}{z_{12}}S' \quad , \quad
T_s(z_1)\bar{S}(z_2)  =  \frac{\frac{3}{2}}{z_{12}^2}\bar{S}+
                         \frac{1}{z_{12}}{\bar S}',  \\
T_s(z_1)T_s(z_2)  =  \frac{3c}{z_{12}^4}+\frac{2}{z_{12}^2}T_s +
                         \frac{1}{z_{12}}T_s' .
\end{array} \right.
\ee
where $z_{12}=z_1-z_2$.

Simple inspection of eqs. (1) and (2) shows that
both the $W_3^{(2)}$ and $N=2$ SCA have the
$Z_2$-grading and the currents with half-integer and integer conformal
spins belong to two different $Z_2$-grading classes. This means, in
particular, that all the currents with integer spins are generated in the
structure relations between the currents with half-integer spins.
Therefore, anyone identification of $J_s, T_s$ with $J_w, T_w$
puts some constraints on the fermionic and half-integer bosonic currents.
We have explicitly checked that these constraints contradict  the Jacobi
identities without introducing  new additional currents with
half-integer spins. So, it is reasonable to expect that
in the desirable minimal case  all the
currents of the $W_3^{(2)}$ and $N=2$ SCA are
independent, i.e. there is no  intersection of these
subalgebras at the embedding in the $N=2$ super-$W_3^{(2)}$ algebra.
Thus the supposed
conformal spins and the number of all the bosonic currents are completely
fixed. They form the $W_3^{(2)}$, Virasoro and $u(1)$ sub-algebras of the
$N=2$ super-$W_3^{(2)}$ algebra.

{\bf 3. The $N=2$ super $W_3^{(2)}$ algebra.}\\
Surprisingly, the advocated algebraic structure has a natural description in
the framework of Polyakov ``soldering'' procedure \cite{p}. In this
approach one writes down a gauge potential $\cal A$ valued in the
(super)algebra of appropriate  (super)group $G$ and performs a ``soldering'' by
putting some components
of $\cal A$ equal to constants. From the residual gauge transformations of
the remaining components of $\cal A$ one can immediately read off the
corresponding OPE, while these components themselves acquire the
interpretation of the currents of algebra.
Here we apply the Polyakov procedure to the case of
supergroup $SL(3|2)$\footnote{An analogous approach with the same gauge
supergroup $SL(3|2)$
was used in ref. \cite{l} with a different "soldering" choice, corresponding
to the case of the $N=2$ super-$W_3$ algebra.}.

Let us start with  the following "soldering" choice for
the $sl(3|2)$-valued gauge potential $\cal A$:
\be \label{3}
{\cal A}=\frac{1}{c} \left(
\begin{array}{ccccc}
2J_s-3J_w & G^{+}     & T_1 & S_1 & S_2 \\
0         & 2J_s-6J_w & G^{-}& 0 & S \\
1 & 0 & 2J_s-3J_w & 0 & S_1 \\
\bar{S}_1 & \bar{S} & \bar{S}_2 & 3J_s-6J_w & T_2 \\
0 & 0 & \bar{S}_1 & 1 & 3J_s-6J_w
\end{array}
\right) ,
\ee
where  $\left\{  J_s,J_w,G^{+},G^{-},T_1,T_2 \right\}$
and $\left\{ S_1,\bar{S}_1,S,\bar{S},S_2,\bar{S}_2 \right\}$
are bosonic and fermionic currents respectively, and will be seen to have the
same spins $\left\{ 1,1,3/2,3/2,2,2 \right\}$.

The potential $\cal A$ obeys the standard infinitesimal gauge transformation
rule
\be \label{var}
\delta {\cal A} = \partial \Lambda + \left[ {\cal A} , \Lambda \right]
\ee
with the $sl(3|2)$-valued matrix of the parameters $\Lambda$
\be \label{par}
\Lambda= \left(
\begin{array}{ccccc}
2l_1+l_2+l_3 & a_1     & a_2 & b_1 & b_2 \\
a_3         & 2l_1-2l_3 & a_4& b_3 & b_4 \\
a_5 & a_6 & 2l_1-l_2+l_3 & b_5 & b_6 \\
c_1 & c_2 & c_3 & 3l_1+l_4 & a_7 \\
c_4 & c_5 & c_6 & a_8 & 3l_1-l_4
\end{array}
\right)  .
\ee
One can find the residual gauge transformations preserving the
form of the gauge potential (\ref{3}). They are transformations with the
following  parameters
\be \label{freepar}
l_1,l_3,a_3,a_5,a_6,a_8,b_3,b_5,(b_1+b_6),c_4,c_5,(c_1+c_6).
\ee
The remaining twelve combinations of the parameters are expressed in terms
of them and the currents. Then one can obtain the transformations of
all the currents in (3) with respect to these constrained gauge
transformations. We
have explicitly checked that if these transformations are represented in the
form
\bea
\delta\phi (z_1) & = & c \oint dz_2 \left[ -6l_1J_s +18l_3J_w+a_3G^{+}
   +a_6G^{-}+3a_5T_1 - 3a_8T_2+b_3\bar{S}+b_5 \bar{S}_2  \right.\nn \\
 & + & \left. (b_1+b_6)\bar{S}_1 - c_4S_2 - c_5S-(c_1+c_6)S_1
       \right] \phi(z_2) ,
\eea
where $\phi(z)$ is any current, the self-consistent OPE for
the currents at the classical level could be obtained.

To make correspondence with the algebras (1) and (2), it is convenient to
perform the following redefinition of the currents:
\be
T_1  =  T_w+\frac{1}{c}S_1\bar{S}_1 -\frac{3}{c}J_w^2  \quad ,\quad
T_2  =  -T_s-\frac{1}{c}S_1\bar{S}_1+\frac{1}{c}J_s^2 .\label{4}
\ee
In terms of the redefined currents we obtained the OPE for $N=2$
super-$W_3^{(2)}$ algebra, which contains, besides the OPE
for the $W_3^{(2)}$ algebra (1) and $N=2$ SCA (2), the following non-trivial
relations:
\bea
& & S_1(z_1)\bar{S}_1(z_2)  =  -\frac{\frac{c}{2}}{z_{12}^2}+
          \frac{\frac{3}{2}J_w-\frac{1}{2}J_s }{z_{12}} \quad, \quad
J_s(z_1)S_1(z_2)  =  -\frac{\frac{1}{2}S_1}{z_{12}} , \nn \\
& & J_w(z_1)S_1(z_2)  =  -\frac{\frac{1}{6}S_1}{z_{12}} \quad, \quad
J_s(z_1)J_w(z_2)  =  \frac{\frac{c}{3}}{z_{12}^2} , \nn \\
& & J_s(z_1)T_w(z_2)  =  \frac{2J_w}{z_{12}^2} \quad, \quad
J_s(z_1)G^{+}(z_2)  =  -\frac{G^{+}}{z_{12}}  , \nn \\
& & J_s(z_1)S_2(z_2)  =  -\frac{\frac{1}{2}S_2}{z_{12}} \quad, \quad
J_w(z_1)T_s(z_2)  =  \frac{\frac{2}{3}J_s}{z_{12}^2} \quad, \quad
J_w(z_1)S_2(z_2)  =  -\frac{\frac{1}{6}S_2}{z_{12}} , \nn \\
& & T_s(z_1)T_w(z_2)  =  \frac{
 \frac{4}{c}\left( S_1\bar{S}_1+J_wJ_s \right)}{z_{12}^2}
                +\frac{
 \frac{2}{c}\left( S_1\bar{S}_2+S_1\bar{S}'_1-S_2\bar{S}_1+
                S_1'\bar{S}_1+2J_wJ_s'\right)}{z_{12}} , \nn \\
& & T_s(z_1)G^{+}(z_2)  =  -\frac{
  \frac{2}{c}\left( G^{+}J_s-S_1\bar{S}\right)}{z_{12}}, \quad
T_s(z_1)S_1(z_2)  =  -\frac{\frac{1}{2}S_1}{z_{12}^2}+
           \frac{S_2-S_1'-
 \frac{1}{c}\left( S_1J_s+3J_wS_1\right)}{2z_{12}} , \nn \\
& & T_s(z_1)S_2(z_2)  =  -\frac{3S_1}{z_{12}^3}+
           \frac{2S_2-3S_1' +\frac{1}{c}\left(
3S_1J_s-9J_wS_1\right)}{2z_{12}^2}
          +  \frac{\frac{2}{c}G^{+}S-\frac{4}{c}S_1T_s+
       \frac{3}{c^2}S_1J_s^2+\frac{1}{c}S_1J_s'}{2z_{12}}  \nn \\
& &       -   \frac{\frac{3}{c}S_2J_s-\frac{3}{c}J_wS_2-\frac{6}{c^2}J_wJ_sS_1+
          \frac{9}{c^2}J_w^2S_1+\frac{6}{c}J_wS_1'-\frac{2}{c}S_1'J_s+
           \frac{3}{c}J_w'S_1-S_2'+S_1''}{2z_{12}} , \nn \\
& & T_w(z_1)S_1(z_2)  =  -\frac{\frac{1}{2}S_1}{z_{12}^2}-
           \frac{S_2-\frac{1}{c}S_1J_s+\frac{5}{c}J_wS_1+S_1'}{2z_{12}}
 \quad, \quad
T_w(z_1)S(z_2)  =
           -\frac{\frac{2}{c}G^{-}S_1-\frac{2}{c}J_wS}{z_{12}} , \nn \\
& & T_w(z_1)S_2(z_2)  =  \frac{3S_1}{z_{12}^3}+
           \frac{2S_2-\frac{3}{c}S_1J_s+\frac{9}{c}J_wS_1+3S_1'}{2z_{12}^2} -
        \frac{\frac{2}{c}G^{+}S+\frac{4}{c}S_1T_w -\frac{1}{c^2}S_1J_s^2+
        \frac{1}{c}S_1J_s'}{2z_{12}}  \nn \\
& &    -  \frac{ \frac{1}{c}S_2J_s-\frac{1}{c}J_wS_2+\frac{6}{c^2}J_wJ_sS_1-
          \frac{21}{c^2}J_w^2S_1-\frac{6}{c}J_wS_1'+\frac{2}{c}S_1'J_s
          -\frac{3}{c}J_w'S_1-S_2'-S_1''}{2z_{12}} , \nn \\
& & G^{+}(z_1)S(z_2)  =  -\frac{2S_1}{z_{12}^2}-
           \frac{S_2-\frac{1}{c}S_1J_s-\frac{3}{c}J_wS_1+S_1'}{z_{12}}
  \quad, \quad
G^{+}(z_1)S_2(z_2)  = -
           \frac{\frac{3}{2c}G^{+}S_1}{z_{12}}  , \nn \\
& & G^{-}(z_1)S_1(z_2)  = -
           \frac{\frac{1}{2}S}{z_{12}} \quad, \quad
G^{-}(z_1)S_2(z_2)  =  \frac{\frac{3}{2}S}{z_{12}^2}-

\frac{\frac{1}{c}G^{-}S_1+\frac{1}{c}J_sS-\frac{9}{c}J_wS-S'}{2z_{12}}
   , \nn \\
& & S_1(z_1)\bar{S}(z_2)  =
           \frac{G^{+}}{2z_{12}} \quad, \quad
S_1(z_1)\bar{S}_2(z_2)  =
           \frac{T_s+T_w+\frac{2}{c}S_1\bar{S}_1-\frac{1}{c}J_s^2-
            \frac{3}{c}J_w^2}{2z_{12}} , \nn \\
& & S(z_1)S_2(z_2)  =
           \frac{3S_1S}{2cz_{12}} \quad, \quad
S(z_1)\bar{S}_2(z_2)  =  -\frac{3G^{-}}{2z_{12}^2}-
           \frac{\frac{3}{c}G^{-}J_s+\frac{1}{c}\bar{S}_1 S-
           \frac{3}{c}J_wG^{-}+{G^{-}}'}{2z_{12}} , \nn \\
& & S_2(z_1)S_2(z_2)  =
           \frac{2S_1S_2}{cz_{12}} ,  \nn \\
& & S_2(z_1)\bar{S}_2(z_2)  =  \frac{3c}{z_{12}^4}+
     \frac{3J_s-9J_w}{z_{12}^3}
    +  \frac{2T_s-2T_w+\frac{1}{c}J_s^2-\frac{18}{c}J_wJ_s+
          \frac{33}{c}J_w^2+3J_s'- 9J_w'}{2z_{12}^2} \nn
\eea
$$
 + \frac{\frac{1}{c}G^{+}G^{-}+\frac{1}{c}S_1\bar{S}_2+
   \frac{1}{c}S\bar{S}+\frac{1}{c}S_2\bar{S}_1
   -\frac{1}{2c^2}J_s^3
  +  \frac{1}{c}T_sJ_s-\frac{1}{c}T_wJ_s-
   \frac{3}{c}J_wT_s+\frac{3}{c}J_wT_w -\frac{3}{2c^2}J_wJ_s^2}{z_{12}}
$$
\be
 +  \frac{\frac{33}{2c^2}J_w^2J_s-\frac{45}{2c^2}J_w^3-
   \frac{9}{2c}J_wJ_s'+\frac{1}{2c}J_s'J_s -\frac{9}{2c}J_w'J_s
 +   \frac{33}{2c}J_w'J_w+\frac{1}{2}T_s'-\frac{1}{2}T_w'
          +\frac{1}{2}J_s''-\frac{3}{2}J_w''}{z_{12}} .
\ee
Here we omitted the OPE that can be obtained from (9)
through the automorphism transformations: $J_{w,s}\rightarrow -J_{w,s}$,
$G^{\pm}\rightarrow \pm G^{\mp}$, $S\rightarrow {\bar S}$,
${\bar S}\rightarrow S$, $S_1\rightarrow {\bar S}_1$,
${\bar S}_1\rightarrow -S_1$,$S_2\rightarrow -{\bar S}_2$,
${\bar S}_2\rightarrow  S_2$.

These OPE are guaranteed to define a closed
nonlinear algebra (with all the Jacobi identities satisfied) because they
have been constructed directly from the gauge
algebra.

Besides the $W_3^{(2)}$ and $N=2$ SCA sub-algebras, the obtained
$N=2$ super-$W_3^{(2)}$ algebra contains the affine Kac-Moody subalgebra
formed by the currents $S_1, \bar{S_1}$ and $3J_w-J_s$.
All the currents with the abovementioned spins
are the primary ones with respect to the following
Virasoro stress-tensor $T$ with a zero central charge :
\be \label{5}
T=T_s+T_w+\frac{4}{c}S_1\bar{S}_1-\frac{4}{c}J_s^2+\frac{12}{c}J_wJ_s-
  \frac{12}{c}J_w^2  \quad ,
\ee
except  $T_s$ and $T_w$ currents which are the quasi primary ones
with  central terms equals to $3c$ and $-3c$ respectively.
We  have explicitly checked that there is no
basis for the $N=2$ super-$W_3^{(2)}$ algebra in which all currents
will be primary  with respect to  any spin 2 currents forming
the Virasoro algebra.

We close this Section with several comments.

First, despite the fact that the constructed $N=2$ super-$W_3^{(2)}$ algebra
has equal numbers of bosonic and fermionic currents, it  seems
impossible to put them in appropriated $N=2$ supermultiplets.
The main obstacle for having the superfield description is the fact that
the numbers of currents with integer and half-integer spins in our
superalgebra do not coincide,
while any $N=2$ superfield comprises an equal number of components with
integer and half-integer spins.

Second, the generalization of the proposed construction
to the case of $N=2$ super-$W_n^{(l)}$ is straightforward like in the case
of $N=2$ super-$W_n$ algebras\cite{l}. Starting
with the gauge potential valued in the superalgebra $sl(n|n-1)$ and
choosing the ``soldering'' that gives rise to
$W_n^{(l)} \times W_{n-1}^{(l-1)}$ algebra in the
bosonic part of $sl(n|n-1)$,
one can read off the corresponding OPE. Of course, there are many
other possibilities for ``soldering'' in the bosonic part of $sl(n|n-1)$.
A detailed discussion of these cases is beyond the scope of this letter.

{\bf 4. Hybrid fields realization.} \\
The $N=2$ super-$W_3^{(2)}$ algebra (1,2,9) has the same property
as  its $W_3^{(2)}$ and $N=2$ SCA
subalgebras, namely, all  currents with integer spins
are generated in the OPE
between the ones with half-integer spins. So, to construct some realization
of the $N=2$ super-$W_3^{(2)}$ algebra, it is sufficient to define only four
spin 3/2 currents $G^{+}, G^{-}, S$ and $\bar{S}$. It also implies that
the $N=2$ super-$W_3^{(2)}$ algebra is a closure of its
$W_3^{(2)}$ and $N=2$ SCA subalgebras with
correlated central charges. Having in mind these useful properties,
let us  construct
hybrid field realizations for $N=2$ super-$W_3^{(2)}$ algebra.

The minimal realization of the $W_3^{(2)}$ and $N=2$ SCA algebras
includes eight basic fields, namely,
the fields $\left\{ {\frac{1}{2}}^B,{\frac{1}{2}}^B,1^B,1^B \right\}$ \cite{b}
and $\left\{ {\frac{1}{2}}^F,{\frac{1}{2}}^F,1^B,1^B \right\}$\cite{z}
respectively. These multiplets are
not sufficient for the construction of the realization of
$N=2$ super-$W_3^{(2)}$ algebra. Let us remind that the
$N=2$ super-$W_3^{(2)}$ algebra
comprises twelve independent currents which
are independent components of the gauge potential $\cal A$ (3)
transforming inhomogeneously (at $c\neq 0$) under
the residual gauge transformations (7).
So, to reproduce these inhomogeneous transformations, it is necessary to
introduce an independent basic field for each current.
For the remaining four
fermionic currents $S_1, \bar{S_1}, S_2$ and $\bar{S_2}$ of the  $N=2$
super-$W_3^{(2)}$ algebra we introduce four fermionic spin 1 basic fields.
Thus, the whole multiplet of basic fields contains  six bosonic fields -
$\left\{ U_1,U_2,V_1,V_2,\xi,\bar\xi \right\}$ and six fermionic ones -
$\left\{ \lambda_1, {\bar\lambda}_1,\lambda_2, {\bar\lambda}_2,\psi,
\bar\psi \right\}$ with the spins $\left\{ 1,1,1,1,\frac{1}{2},\frac{1}{2}
\right\}$, respectively, and with the  $J_s$- and $J_w$-charges equal to
the charges of corresponding currents.

With all the above knowledge, we are ready to construct the hybrid  fields
realization.
Taking the most general ansatz for the
currents (in terms of the introduced basic fields) as well as  for the
OPE between basic fields  and
requiring the consistency with the OPE (1,2,9) we obtain the following
realization
\bea
S & = &  \sqrt{c}\left( \psi'-\frac{3}{2}V_1\psi \right)
      - {\bar\xi}\bar\lambda_2 +
     \frac{1}{\sqrt{c}}\left( \xi\bar\xi\psi + U_1\psi+U_2\psi \right), \nn \\
{\bar S} & = & \sqrt{c}\left( \bar\psi'-\frac{1}{2}V_2\bar\psi\right)
     + \xi\lambda_1
     -\frac{1}{\sqrt{c}}\left( \xi\bar\xi\bar\psi- U_1\bar\psi \right), \nn \\
G^{+} & = & \sqrt{c}\left( V_1\xi - \frac{1}{2}V_2\xi +\xi' \right)
  + \bar\lambda_1\bar\psi
  +\frac{1}{\sqrt{c}}\left( U_1\xi -\xi\xi\bar\xi -
   \xi\psi\bar\psi  \right), \nn \\
G^{-} & = & -\sqrt{c}\left( \frac{3}{2}V_1\bar\xi + V_2\bar\xi
           - \bar\xi' \right) + \lambda_2\psi
  +\frac{1}{\sqrt{c}}\left( \xi\bar\xi\bar\xi +
   \bar\xi\psi\bar\psi + U_1\bar\xi + U_2\bar\xi \right), \nn \\
J_s & = & \frac{c}{4}\left( 3V_1-V_2 \right)
        -\frac{1}{2}U_2-\xi\bar\xi-
       \frac{1}{2}\psi\bar\psi ,\nn \\
J_w & = & \frac{c}{12}\left( 5V_1+V_2 \right) -\frac{1}{6}U_2-
    \frac{1}{2}\xi\bar\xi-\frac{1}{3}\psi\bar\psi , \nn \\
S_1 & = & \frac{\sqrt{c}}{2}\left( \bar\lambda_1 + \bar\lambda_2\right) -
        \frac{1}{2}\xi\psi ,\nn \\
\bar{S}_1 & = & \frac{\sqrt{c}}{2}\left( \lambda_1 + \lambda_2 \right) -
        \frac{1}{2}\bar\xi\bar\psi ,\nn \\
S_2 & = & \frac{\sqrt{c}}{2}\left( V_2\bar\lambda_2 -V_1\bar\lambda_1
  + \bar\lambda_1'-\bar\lambda_2'\right)
  +\frac{3}{2}V_1\xi\psi- \frac{1}{2}\xi\psi'-\frac{1}{2}V_2\xi\psi
  +\frac{1}{2}\xi'\psi \nn \\
  & - & \frac{1}{\sqrt{c}}\left(
       \frac{1}{4}\xi\bar\xi\bar\lambda_1-
        \frac{3}{4}\xi\bar\xi\bar\lambda_2+
        \frac{3}{4}\psi\bar\psi\bar\lambda_1-
        \frac{1}{4}\psi\bar\psi\bar\lambda_2-
        U_1\bar\lambda_1
       +U_1\bar\lambda_2
       -\frac{1}{2}U_2\bar\lambda_1
       +\frac{1}{2}U_2\bar\lambda_2 \right) \nn \\
    &- & \frac{1}{2c}\left( \frac{3}{2}\xi\xi\bar\xi\psi+ U_2\xi\psi\right)
,\nn \\
\bar{S}_2 & = & -\sqrt{c}\left(  \frac{1}{2}V_1\lambda_1-
           V_1\lambda_2 + V_2\lambda_1-\frac{1}{2}V_2\lambda_2
          -\frac{1}{2}\lambda_1'+\frac{1}{2}\lambda_2' \right)
          + \frac{1}{2}\bar\xi\bar\psi'-\frac{1}{2}\bar\xi'\bar\psi
        + \frac{1}{2}V_1\bar\xi\bar\psi
        +\frac{1}{2}V_2\bar\xi\bar\psi \nn \\
    &+ & \frac{1}{\sqrt{c}}\left(
        \frac{3}{4}\xi\bar\xi\lambda_1-
       \frac{1}{4}\xi\bar\xi\lambda_2+
       \frac{1}{4}\psi\bar\psi\lambda_1-
       \frac{3}{4}\psi\bar\psi\lambda_2
       +U_1\lambda_1
      -U_1\lambda_2+\frac{1}{2}U_2\lambda_1
       - \frac{1}{2}U_2\lambda_2 \right)  \nn \\
    &- & \frac{1}{c}\left( \frac{3}{4}\xi\bar\xi\bar\xi\bar\psi
       + \frac{1}{2}U_2\bar\xi\bar\psi \right) ,\nn \\
T_s & = & \frac{c}{4}\left( 3V_1' +V_2' -3V_1V_2\right)+
          \lambda_1\bar\lambda_2+
         \frac{1}{2}\psi\bar\psi'-\frac{1}{2}\psi'\bar\psi
     -\frac{3}{2}V_1\xi\bar\xi+
          \frac{1}{2}V_2\xi\bar\xi +\frac{3}{2}U_1V_1 +\frac{1}{2}U_1V_2\nn \\
     &+ & \frac{1}{2}U_2V_2 -U_1'-\frac{1}{2}U_2'
      -\frac{1}{\sqrt{c}}\left(
\xi\lambda_1\psi-\bar\xi\bar\lambda_2\bar\psi\right)
          +\frac{1}{c}\left( U_2\xi\bar\xi
         -U_1U_1-U_1U_2
         +\xi\xi\bar\xi\bar\xi\right) ,\nn \\
T_w & = & \frac{c}{4}\left(
         \frac{7}{3}V_1V_1 +\frac{7}{3}V_1V_2+ \frac{7}{3}V_2V_2
          -V_1' -3V_2'\right)
         -\bar\lambda_1\lambda_2
        - \frac{1}{2}\xi\bar\xi'+
         \frac{1}{2}\xi'\bar\xi \nn \\
    &-&  \frac{5}{6}V_1\psi\bar\psi
       - \frac{1}{6}V_2\psi\bar\psi-\frac{1}{2}U_1V_1-\frac{3}{2}U_1V_2
        -\frac{2}{3}U_2V_1- \frac{5}{6}U_2V_2+U_1'+\frac{1}{2}U_2' \nn \\
    &-&  \frac{1}{\sqrt{c}}\left( \xi\lambda_2\psi-
         \bar\xi\bar\lambda_1\bar\psi \right)
           +  \frac{1}{c}\left( U_1U_1+U_1U_2
          +\frac{1}{3}U_2U_2 + \frac{1}{3}U_2\psi\bar\psi \right) .
\eea
The basic fields form the following superalgebra
\bea
& & \xi(z_1)\bar\xi(z_2) =  -\frac{1}{z_{12}} , \quad
\psi(z_1)\bar\psi(z_2)  =  -\frac{1}{z_{12}} , \quad
\lambda_1(z_1)\bar\lambda_1(z_2)  =  \frac{1}{z_{12}^2}+\frac{V_1}{z_{12}}, \nn
\\
& & \lambda_2(z_1)\bar\lambda_2(z_2)  =
   \frac{1}{z_{12}^2}+\frac{V_2}{z_{12}}, \quad
U_2(z_1)V_2(z_2)  =  -\frac{1}{z_{12}^2}, \quad
U_2(z_1)V_1(z_2)  =  -\frac{1}{z_{12}^2}, \nn \\
& & U_1(z_1)V_1(z_2)  =  \frac{1}{z_{12}^2}, \quad
U_1(z_1)\lambda_1(z_2)  =  \frac{\lambda_1}{z_{12}}, \quad
U_1(z_1)\bar\lambda_1(z_2)  =  -\frac{\bar\lambda_1}{z_{12}}, \quad
U_2(z_1)\lambda_1(z_2)  =  -\frac{\lambda_1}{z_{12}} ,\nn \\
& & U_2(z_1)\bar\lambda_1(z_2)  =  \frac{\bar\lambda_1}{z_{12}}, \quad
U_2(z_1)\lambda_2(z_2)  =  -\frac{\lambda_2}{z_{12}},\quad
U_2(z_1)\bar\lambda_2(z_2)  =  \frac{\bar\lambda_2}{z_{12}} .
\eea
It is instructive to examine the structure of the
stress-tensor $T$ (10) in this realization
\be
T=-\lambda_1\bar\lambda_1-\lambda_2\bar\lambda_2 +\frac{1}{2}\xi'\bar\xi-
   \frac{1}{2}\xi\bar\xi'+\frac{1}{2}\psi\bar\psi'-\frac{1}{2}\psi'\bar\psi
  +U_1V_1-U_1V_2-U_2V_2+\frac{c}{2}V_1'-\frac{c}{2}V_2' \; .
\ee
As we would expect,  it is quadratic in all basic fields unlike
the $T_w$ and $T_s$ in (11). The
obtained relations (11) are defined up to possible automorphisms of both
the $N=2$ super-$W_3^{(2)}$ algebra and  the basic algebra (12). In
particular, the OPE (12)  possesses the one-parametric automorphism
transformations
\be
U_1(x)\rightarrow \tilde{U}_1(x) = U_1(x) - \alpha V_2(x) \quad , \quad
U_2(x)\rightarrow \tilde{U}_2(x) = U_2(x) + \alpha \left( V_2(x)-V_1(x) \right)
,
\ee
where $\alpha$ is an arbitrary constant.

{\bf 5. Contraction $c \rightarrow 0$.} \\
Let us discuss the $c\rightarrow 0$ contraction
of our $N=2$ super-$W_3^{(2)}$ algebra.
We will demonstrate that one can find
different non isomorphic nonlinear versions of the $N=2$ super-$W_3^{(2)}$
algebra in the contraction limit $c=0$. The nature of this phenomenon is
connected with the $1/c$-singularities of structural constants in the
OPE (1,2,9). However, the structure of these singularities is basis-
dependent. There are, of course, many different possible choices of the
basis for the superalgebra we have constructed and, in principle,
any contraction could be analyzed in  terms of OPE. Nevertheless,
it is rather easy to performe that analysis on the level of given
realization (11).

One can see, that the obtained realization (11) has
singularities as $c\rightarrow 0$ too, which is simply a reflection of the
abovementioned facts. The coefficients of different independent composite
structures have  different orders of the
$1/c$-singularities as $c\rightarrow 0$. To approach the limit $c=0$,
one should rescale the basis currents for elimination of all the leading
singularities as $c\rightarrow 0$.
The bases with  different structures of the
leading $1/c$-singularities lead to  different nonisomorphic algebras
(after rescaling and taking the limit $c\rightarrow 0$).
At least two different superalgebras at
$c=0$  can be obtained with the help of the  realization (11). For this
purpose let us consider two basises. The first one contains the same
currents as in (1,2,9) while the second one contains the same currents
as the first one except $T_w$ which is replaced by $T$ (10,13). The $T$ is
regular unlike $T_w$ which contains the $1/c$-singularities. Rescaling the
currents in accordance with the structure of their leading
$1/c$-singularities as
\bea
& & \tilde{J_w}  =  J_w,\quad
\tilde{J_s}  =  J_s ,\quad
\tilde{G^{+}}  =  \sqrt{c}G^{+},\quad
\tilde{G^{-}}  =  \sqrt{c}G^{-},\quad
\tilde{T_w}  =  cT_w, \nn \\
& & \tilde{T_s}  =  cT_s ,\quad
\tilde{T}  =  T ,\quad
\tilde{S}  =  \sqrt{c} S,\quad
\widetilde{\bar S}  =  \sqrt{c}{\bar S}, \nn \\
& & \tilde{S_1}  =   S_1  ,\quad
\widetilde{\bar S_1}  =  {\bar S_1} ,\quad
\tilde{S_2}  =  cS_2  ,\quad
\widetilde{\bar{S}_2}  =  c\bar{S}_2 .
\eea
and taking the contraction
limit $c=0$, one can get the OPE for two different superalgebras.
We do not adduce these rather complicate expressions here because
they can be obtained straightforwardly from (1,2,9,15).

{\bf 6. Conclusion}\\
To summarize, we have constructed the classical $N=2$ super-$W_3^{(2)}$
algebra and its hybrid field realization.
We have also found the $N=2$ classical nonlinear superalgebras
at $c=0$. The generalization to the case of $N=2$ super-$W_n^{(l)}$
algebras is discussed.

    In a forthcoming publication \cite{ahn} we will extend our
consideration to the  full quantum $N=2$ super-$W_3^{(2)}$ algebra.
We are planning also to consider the general hamiltonian
flows on the $N=2$ super-$W_3^{(2)}$ algebra and the
corresponding equations  which have a great chance to be integrable.

{\bf Acknowledgments}.\\
We would like to thank  C.Ahn, S.Bellucci, A.Pashnev and
especially E.Ivanov for many
useful and clarifying discussions and V.Ogievetsky for his interest in this
work.
We are also indebted to K.Thielemans
because without his Mathematica\cite{Math} packet OPEdefs\cite{Kris} many
cumbersome calculations would spent much more time.
One of us (S.K.) wishes to thank
the Laboratori Nazionali di Frascati for the hospitality during
the course of this work.

\end{document}